\documentclass[10pt,twocolumn,twoside]{article}

\usepackage{palatino}  
\usepackage[T1]{fontenc}

\usepackage[left=3.5cm,top=2.54cm,right=2.5cm,bottom=2.8cm,nohead]{geometry} 
\usepackage{amsmath}
\usepackage[font=sf]{caption}
\usepackage{color}
\usepackage{fancyhdr}
\usepackage{graphicx}
\usepackage{lettrine}
\usepackage[super,sort&compress]{natbib}
\usepackage{pifont}
\usepackage{sectsty}
\usepackage{setspace}
\usepackage{sidecap}  
\usepackage{subfig}
\usepackage{wrapfig}

\definecolor{sectcol}{rgb}{0.56,0.0,0.33} 
\definecolor{dropped}{rgb}{0.56,0.0,0.33}
\definecolor{update}{rgb}{0.098,0.357,0.675} 
\definecolor{lightgray}{rgb}{0.95,0.95,0.95}
\sectionfont{\large\sffamily\color{sectcol}\vspace{-2mm}}
\subsectionfont{\normalsize\sffamily\bfseries\vspace{-2mm}}

\newlength{\up}
\usepackage[
    pdftex,
    pdftitle={},
    pdfauthor={L.~A. Barba},
    pdfpagemode={UseOutlines},
    bookmarks, bookmarksopen,bookmarksnumbered={True},
    colorlinks, linkcolor={black},citecolor={black},urlcolor={black}
    ]{hyperref}

\begin{document}
\pagestyle{fancy}

\twocolumn[ 
{ \sf {\LARGE{\color{sectcol}Reproducible and replicable CFD: it's harder than you think}}
}
\vspace{0.8cm}

{ \sf 
\textbf {Olivier Mesnard, Lorena A. Barba}\\
  Mechanical and Aerospace Engineering, George Washington University, Washington DC 20052
}

\vspace{1.0cm}

{ \sf 
{ \textit{Completing a full replication study of our previously published findings
on bluff-body aerodynamics was harder than we thought. Despite the fact
that we have good reproducible-research practices, sharing our code and
data openly. Here's what we learned from three years, four CFD codes and
hundreds of runs.}}
}
\vspace{1.5cm}

] 

\lettrine{\textcolor{dropped}{O}}{}ur research group prides itself for having adopted Reproducible
Research practices. Barba (2012)\cite{barba2012} made a public pledge titled
\emph{``Reproducibility PI Manifesto''} (PI: Principal Investigator),
which at the core is a promise to make all research materials and
methods open access and discoverable: releasing code, data and
analysis/visualization scripts.

In 2014, we published a study on Physics of Fluids titled \emph{``Lift
and wakes of flying snakes''}.\cite{krishnan2014} It is a study that
used our in-house code for solving the equations of fluid motion in two
dimensions (2D), with a solution approach called the ``immersed boundary
method.'' The key of such a method for solving the equations is that it
exchanges complexity in the mesh generation step for complexity in the
application of boundary conditions. It makes it possible to use a simple
mesh for discretization (structured Cartesian), but at the cost of an
elaborate process that interpolates values of fluid velocity at the
boundary points to ensure the no-slip boundary condition (that fluid
sticks to a wall). The main finding of our study on wakes of flying
snakes was that the 2D section with anatomically correct geometry for
the snake's body experiences lift enhancement at a given angle of
attack. A previous experimental study\cite{HoldenEtAl2014} had already
shown that the lift coefficient of a snake cross section in a wind
tunnel gets an extra oomph of lift at 35 degrees angle-of-attack. Our
simulations showed the same feature in the plot of lift coefficient.\cite{krishnan2013}
Many detailed observations of the wake (visualized
from the fluid-flow solution in terms of the vorticity field in space
and time) allowed us to give an explanation of the mechanism providing
extra lift. It arises from a vortex on the dorsal side of the body
remaining closer to the surface under the effects of interactions with
secondary vorticity. The flow around the snake's body cross section
adopts a pattern known as a von Karman vortex street. It is a
particularly complex flow, because it involves three shear layers: the
boundary layer, a separating free shear layer, and the wake.\cite{Williamson_1996}
Physically, each of these shear layers is
subject to instabilities. The free shear layer can experience 2D
Kelvin-Helmholtz instability, while the wake experiences both 2D and 3D
instabilities and can show chaotic behavior. Such flows are particularly
challenging for computational fluid dynamics (CFD).

When a computational research group produces this kind of study with an
in-house code, it can take one, two or even three years to write a full
research software from scratch, and complete verification and
validation. Often, one gets the question: why not use a commercial CFD
package? Why not use another research group's open-source code? Doesn't
it take much longer to write yet another CFD solver than to use existing
code? Beyond reasons that have to do with inventing new methods, it's a
good question. To explore using an existing CFD solver for future
research, we decided to first complete a full replication of our
previous results with these alternatives. Our commitment to open-source
software for research is unwavering, which rules out commercial
packages. Perhaps the most well known open-source fluid-flow software is
OpenFOAM, so we set out to replicate our published results with this
code. A more specialist open-source code is IBAMR, a project born at New
York University that has continued development for a decade. And
finally, our own group developed a new code, implementing the same
solution method we had before, but providing parallel computing via the
renowned PETSc library. We embarked on a full replication study of our
previous work, using three new fluid-flow codes.

This is the story of what happened next: three years of dedicated work
that encountered a dozen ways that things can go wrong, conquered one
after another, to arrive finally at (approximately) the same findings
and a whole new understanding of what it means to do ``reproducible
research'' in computational fluid dynamics.

\begin{figure*}
    \colorbox{lightgray}{
     \begin{minipage}[c]{0.98\textwidth}
      \bigskip
\small \sffamily{
\subsection*{Fluid-flow solvers we used:}
{\sf \footnotesize
\textbf{cuIBM}--- Used for our original study (Krishan et al., 2014),
this code is written in C CUDA to exploit GPU hardware, but is serial on
CPU. It uses the NVIDIA \emph{Cusp} library for solving sparse linear
systems on GPU. \url{https://github.com/barbagroup/cuIBM}\\
\textbf{OpenFOAM}--- A free and open-source CFD package that includes a
suite of numerical solvers. The core discretization scheme is a
finite-volume method applied on mesh cells of arbitrary shape.
\url{http://www.openfoam.org}\\
\textbf{IBAMR}--- A parallel code using the immersed boundary method on
Cartesian meshes, with adaptive mesh refinement.
\url{https://github.com/ibamr/ibamr}\\
\textbf{PetIBM}--- Our own re-implementation of \emph{cuIBM}, but for
distributed-memory parallel systems. It uses the PETSc library for
solving sparse linear systems in parallel.
\url{https://github.com/barbagroup/PetIBM} 
}}
\vspace{0.4cm}
\end{minipage}      
}
\end{figure*}

\section*{Story 1: Meshing and boundary conditions can ruin
everything}\label{story-1-meshing-and-boundary-conditions-can-ruin-everything}

Generating good meshes for discretization is probably the most vexing
chore of computational fluid dynamics. And stipulating boundary
conditions on the edge of a mesh takes some nerve, too. An early example
of how frustrating it can be to investigate different outflow boundary
conditions is reported in Sani et al.\ (1994).\cite{SaniGresho1994} Our first attempts at
a full replication study of the 2D snake aerodynamics with IcoFOAM, the
incompressible laminar Navier-Stokes solver of OpenFOAM, showed us just
how vexing and unnerving these issues can be.

OpenFOAM can take various types of mesh as input. One popular mesh
generator is called GMSH: it produces triangles that are as fine as you
want them near the body, while getting coarser as the mesh points are
farther away. Already, we encounter a problem: how to create a mesh of
triangles that gives a comparable resolution to that obtained with our
original structured Cartesian mesh? After dedicated effort, we produced
the best mesh we could that matches our previous study in the finest
cell width near the body. But when using this mesh to solve the fluid
flow around the snake geometry, we got spurious specks of high vorticity
in places where there shouldn't be any (Figure 1). Even though the
meshes passed the OpenFOAM quality checks, these unphysical vortices
appeared for any flow Reynolds number or body angle of attack we
tried---although they were not responsible for the simulations to blow
up (fail due to rapid error growth). Finally, we gave up with the
(popular) GMSH and tried another mesh generator: SnappyHexMesh (details
and plots of the meshes are included in the supplementary materials).
Success! No unphysical patches in the vorticity field this time. But
another problem persisted: after the wake vortices hit the edge of the
computational domain in the downstream side, a nasty back pressure
appeared there and started propagating to the inside of the domain
(Figure 2). This situation is also unphysical, and we were certain there
was a problem with the chosen outflow boundary condition in OpenFOAM,
but did not find any way to stipulate another, more appropriate boundary
condition. We used a zero-gradient condition for the pressure at the
outlet (and tried several other possibilities), which we found was a
widespread choice in the examples and documentation of OpenFOAM. After
months, one typing mistake when launching a run from the command line
made OpenFOAM print out the set of available boundary conditions, and we
found that an \emph{advective} condition was available that could solve
our problem. All this time, we were looking for a \emph{convective}
condition, which is just another name for the same thing: satisfying a
linear convection equation at the boundary points. Finally, simulations
with OpenFOAM were looking correct---and happily, the main feature of
the aerodynamics was replicated: an enhanced lift coefficient at 35 degrees angle-of-attack (Figure 3). But not all is perfect. The time signatures
of lift and drag coefficient do show differences between our IcoFOAM
calculation and the original published ones (Figure 4). The key finding
uses an \emph{average} lift coefficient, calculated with data in a time
range that is reasonable but arbitrary. Refining the mesh or reducing
the exit criterion of the iterative solvers made a difference of less
than 0.5\% in this quantity. The average force coefficients match
(within \textless{} 3\%) our previous results, despite the differences
seen on the time series. Are these the same solutions? Is it acceptable
as a replication study? We think yes, but this is a judgement call.

\textbf{Postmortem}. IcoFOAM solves the fluid equations using a
finite-volume method in an unstructured grid, while our published study
used an immersed boundary method in a stretched Cartesian grid.
Comparing results obtained under such different conditions is a delicate
operation. We made our best attempt at creating a fluid mesh for
OpenFOAM that was of similar resolution near the body as we had used
before. But unstructured grids are complex geometrical objects. Two
unstructured meshes built with the same parameters will not be exactly
the same, even. The mesh-generation procedures are not necessarily
deterministic, and regularly produce bad triangles that need to be
repaired. The complications of building a \emph{good quality} mesh is
one of the reasons some prefer immersed boundary methods!

\begin{figure}
\begin{center}
\includegraphics[width=0.98\columnwidth]{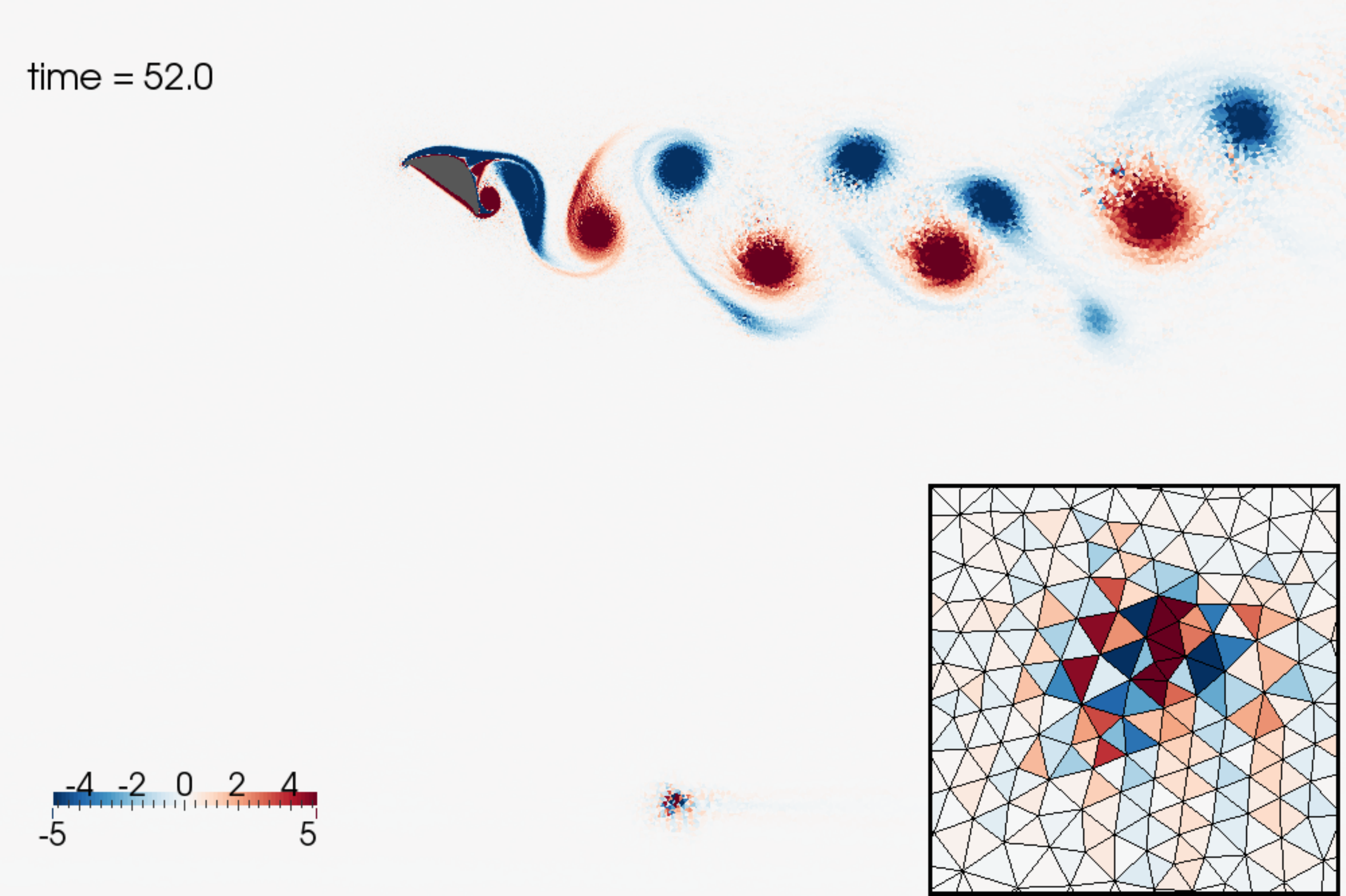}
\caption{{Vorticity field after 52 time-units of flow-simulation with IcoFOAM for
a snake's section with angle-of-attack 35 degrees and Reynolds number
2000. We created a triangular mesh (about 700k triangles) with the free
software GMSH. The box insert at the bottom-right shows a zoom-in to the
portion of the mesh with spurious vorticity (seen bottom-center of the
main plot). \label{figure1}%
}}
\end{center}
\end{figure}

\begin{figure}
\begin{center}
\includegraphics[width=0.98\columnwidth]{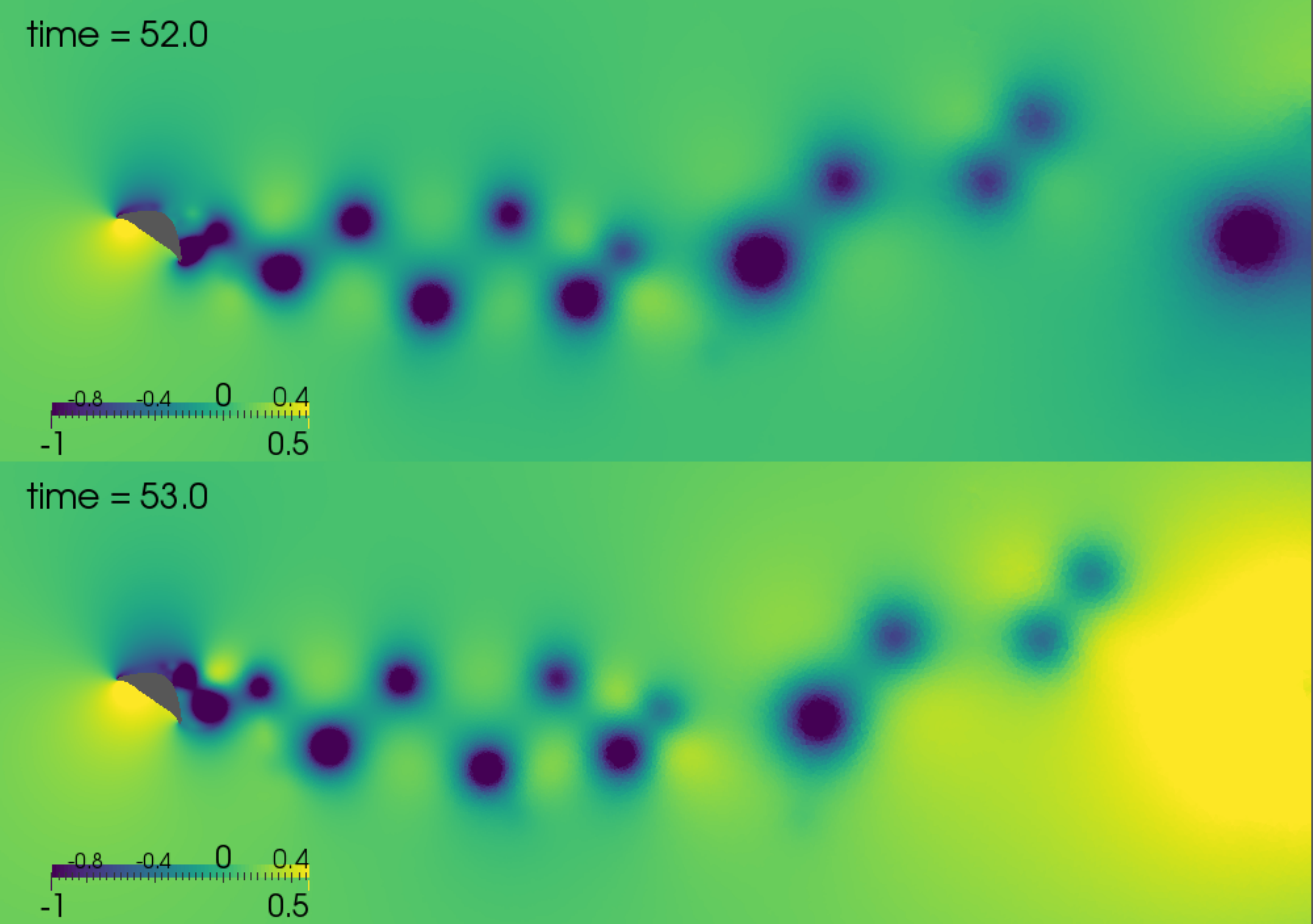}
\caption{{Pressure field after 52 (top) and 53 (bottom) time-units of
flow-simulation with IcoFOAM for snake section with angle-of-attack 35
degrees and Reynolds number 2000. The simulation crashed after about 62
time-units because of the back pressure at the outlet boundary.
\label{figure2}%
}}
\end{center}
\end{figure}

\begin{figure}
\begin{center}
\includegraphics[width=0.9\columnwidth]{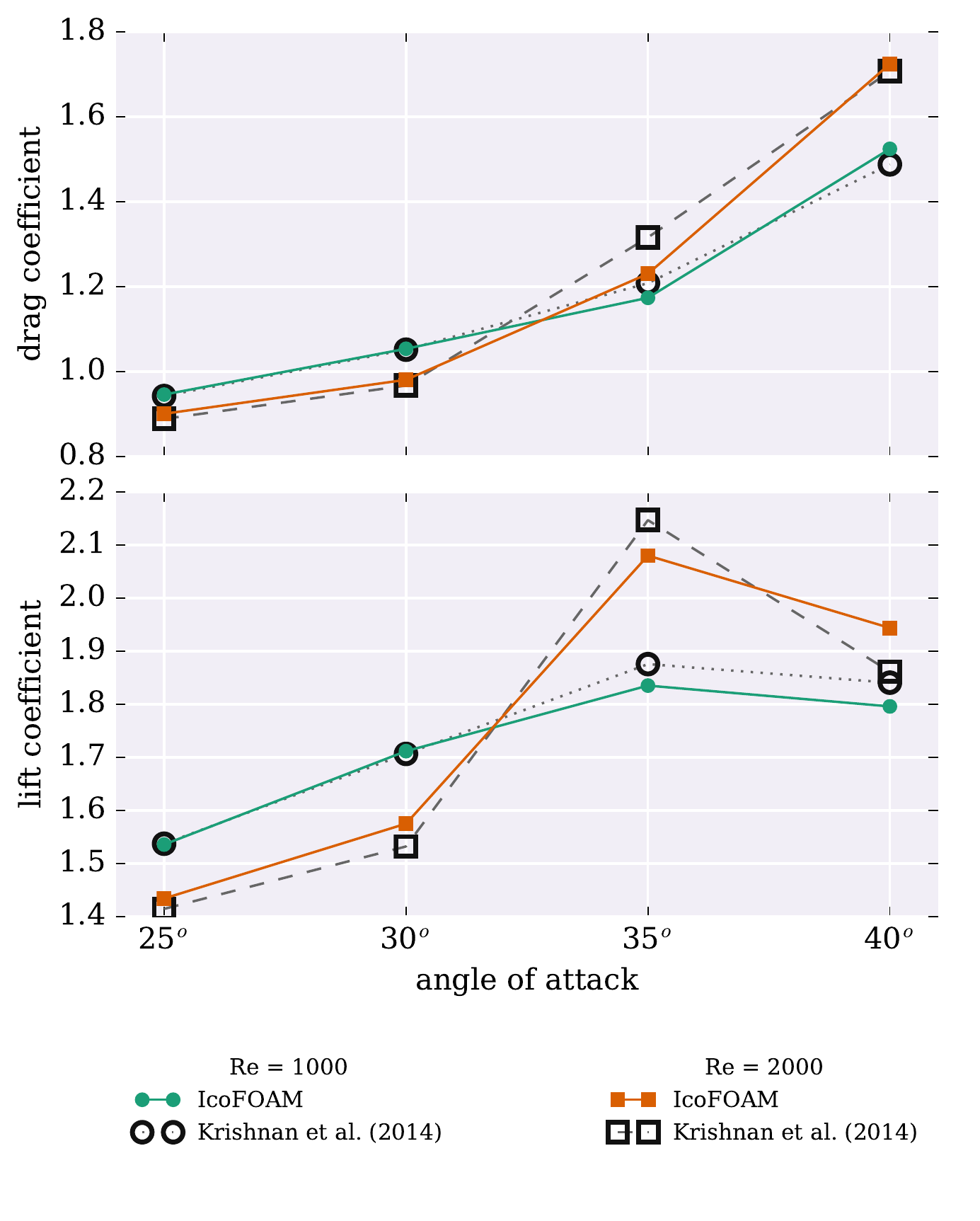}
\caption{{Time-averaged drag (top) and lift (bottom) coefficients as function of
the snake's angle-of-attack for Reynolds numbers 1000 and 2000. We
averaged all IcoFOAM force coefficients between 32 and 64 time-units of
flow-simulation as we have done in our previous study. \label{figure3}%
}}
\end{center}
\end{figure}

\begin{figure}[t]
\begin{center}
\includegraphics[width=0.9\columnwidth]{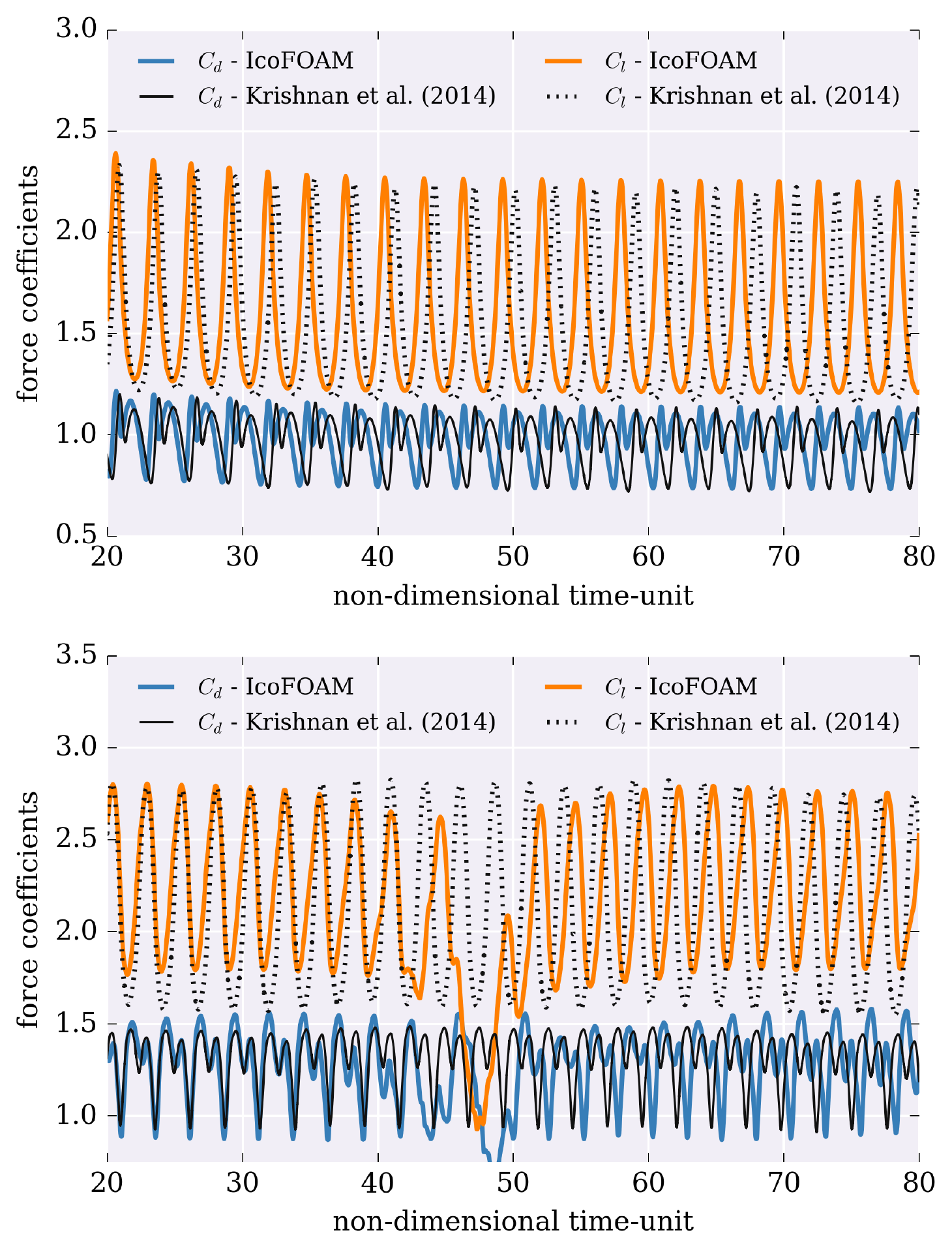}
\caption{{Instantaneous force coefficients on the snake's section with
angle-of-attack 30 degrees (top) and 35 degrees (bottom) at Reynolds
number 2000. We compare the IcoFOAM results with the cuIBM results from
our previous study. We created a 3.4 million cells (mostly hexahedra)
with SnappyHexMesh, one of the OpenFOAM mesh utilities. \label{figure4}%
}}
\end{center}
\end{figure}

\section*{Story 2: You can hit snags with other researchers'
codes}\label{story-2-you-can-hit-snags-with-other-researchers-codes}

Open-source research software can often be poorly documented and
unsupported, and on occasion it can even be an unreadable mess. But in
this case, we are in luck. IBAMR is a solid piece of software, the code
is documented, and you can even get swift response from the authors via
the topical online forum. The developers don't provide a user's manual,
but they have plenty of examples within the code repository. Still,
mastering other researchers' code is challenging and we hit a couple of
snags that complicated the journey.

IBAMR is described as ``an adaptive and distributed-memory parallel
implementation of the immersed boundary method.'' The essence of the
immersed boundary method is that the fluid is represented by a
structured mesh, while the solid boundary is represented by its own,
separate mesh that moves with the body. We speak of an Eulerian mesh for
the fluid, and a Lagrangian mesh for the solid. The forces exerted by
the fluid on the body, and vice versa, appear as an additional integral
equation and interpolation schemes between the two meshes. The role of
these is to make the fluid ``stick'' to the wall (no-slip boundary
condition) and allow the body to feel aerodynamic forces (lift and
drag). Our cuIBM code uses a variant called the immersed-boundary
projection method.\cite{taira2007} IBAMR is a library that provides
different methods,\cite{bhalla2013} but despite the variations, we
assumed it would work similarly.

We already know that boundary conditions at the outlet of the
computational domain can be problematic. This is no different with
immersed boundary methods. Our first attempt with IBAMR used a boundary
condition at the outlet following their example for flow around a
circular cylinder (this turned out to be a traction-free boundary
condition). Unfortunately, it resulted in a spurious blockage of the
wake vortices when they reach the domain boundary: strong vorticity
rebounded from the artificial boundary and propagated back to the domain
(Figure 5, top). Of course, this is unphysical and the result is
unacceptable.

In a conversation with the main developers on the online forum, they
suggested a work-around: using ``boundary stabilization,'' which adds a
forcing to push the vortices out. (IBAMR does not yet provide a
convective/advective boundary condition.) With this new configuration,
the simulations of the snake profile resulted in a wake that looked
physical (Figure 5, bottom), but a computed lift coefficient that was
considerably different from our published study (Figure 6). Another dive
into Bhalla et al. (2015) led us to notice that the benchmark examples
were set up in a way unexpected to us: the no-slip condition is forced
\emph{inside} the body, and not just on the boundary. Immersed boundary
methods normally apply the no-slip constraint on boundary points only.
When we followed their examples, our simulations with IBAMR were able to
replicate the lift enhancement at 35 degrees angle-of-attack, although
with a slightly different value of average lift (\textless{} 5\% off).
If we look at the time signature of the lift and drag coefficients,
there is excellent agreement with our previous results for 30 degrees
angle-of-attack (Re=2000). But at 35 degrees, the time signatures drift
apart after about 40 time units (more than 150 thousand time steps).
There is a marked drop in the (time varying) lift coefficient (Figure
7), but because the average is calculated over a time range between 32
and 64 time units (a reasonable but arbitrary choice), the final numeric
result is not far off our published study. To start, we matched the mesh
resolution in the vicinity of the body. Refining the mesh further,
reducing the exit criterion of the iterative solver, or enlarging the
computational domain did not improve things. Reducing the time
increment, however, did. In Figure 7, we show the time-varying lift
coefficients obtained with the parameter CFL set to 0.3 and 0.1---the
CFL, or Courant-Friedrichs-Lewy number, constrains the ratio of time
increment to grid spacing. Like in the previous case, using OpenFOAM, we
make a judgement call that our result with IBAMR does indeed pass muster
as a replication of our previous study.

\textbf{Postmortem}. Even the best open-source research code can have
unexpected attributes that only the original authors know in depth. We
still don't understand \emph{why} IBAMR requires interior body points to
be constrained, despite insistent reading of the literature. One issue
that affects our community is that we don't expect authors to provide in
their papers all the details, nor do we require papers to be accompanied
by code and data. We learned from this experience that using an open
research code and getting correct results with it could involve a long
investigative period, potentially requiring communication with the
original authors and many failed attempts. If the code is not documented
and the original authors not responsive to questions, then building your
own code from scratch could be more sensible!

\begin{figure}
\begin{center}
\includegraphics[width=0.9\columnwidth]{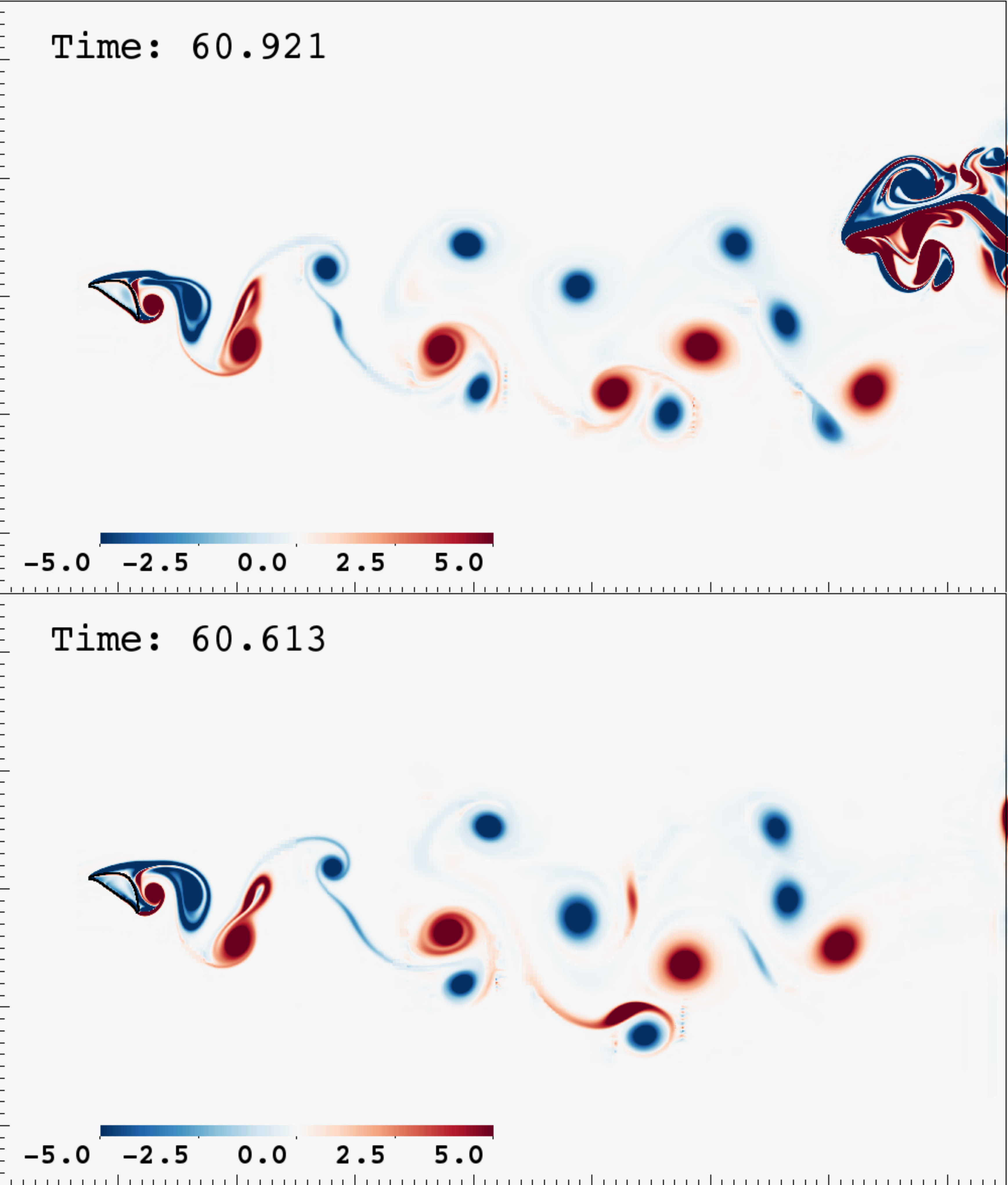}
\caption{{Vorticity field after about 61 time-units of flow-simulation with IBAMR
for a snake's section with angle-of-attack 35 degrees and Reynolds
number 2000. Top: without boundary stabilization at the outlet; bottom:
with boundary stabilization. \label{figure5}%
}}
\end{center}
\end{figure}

\begin{figure}
\begin{center}
\includegraphics[width=0.9\columnwidth]{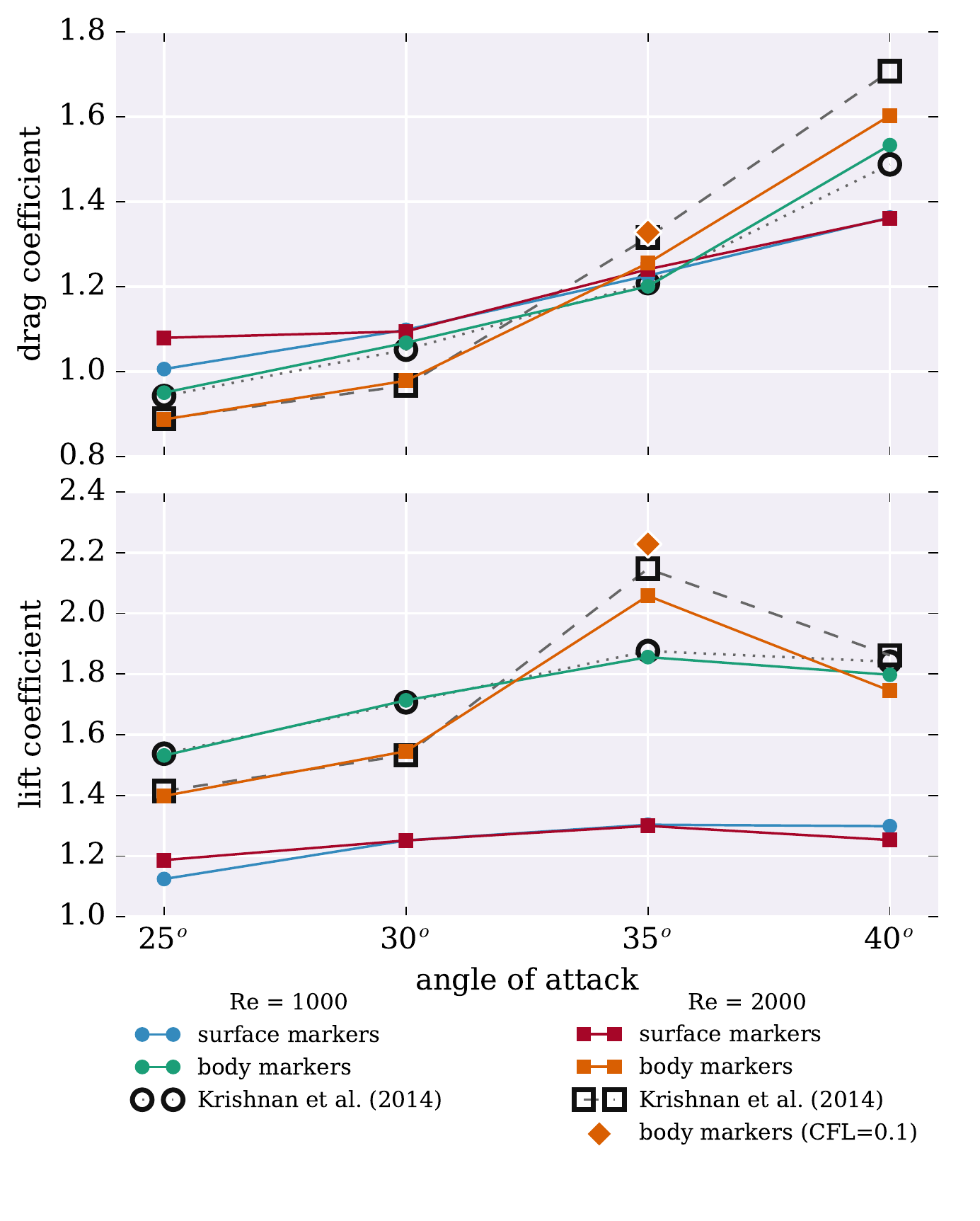}
\caption{{Time-averaged drag (top) and lift (bottom) coefficients as function of
the angle-of-attack of the snake's section for Reynolds numbers 1000 and
2000. We averaged each force signal between 32 and 64 time-units of
flow-simulation with IBAMR to compare with our previous results.
\label{figure6}%
}}
\end{center}
\end{figure}

\begin{figure}
\begin{center}
\includegraphics[width=0.9\columnwidth]{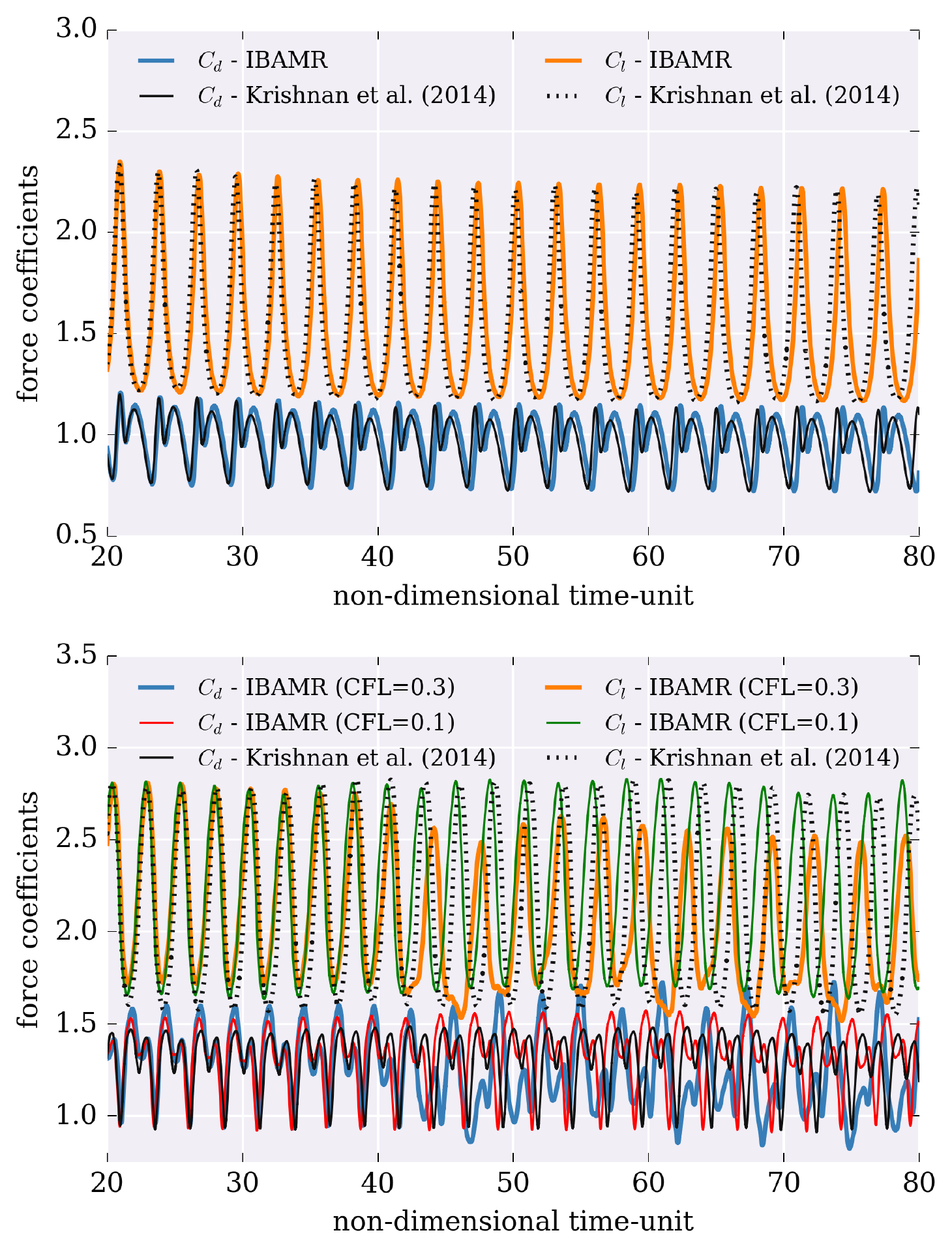}
\caption{{Instantaneous force coefficients at Reynolds number 2000 for the snake's
section at angle-of-attack 30 (top) and 35 (bottom) degrees. Here, the
no-slip condition is enforced inside the section. We compare the IBAMR
results with cuIBM ones from our past study. \label{figure7}%
}}
\end{center}
\end{figure}

\section*{Story 3: All linear algebra libraries are not created
equal}\label{story-3-all-linear-algebra-libraries-are-not-created-equal}

Our previous study used cuIBM, running on a single GPU device. The
largest problem that we can fit in the memory of a high-end GPU has just
a few million mesh points, which is not enough to solve
three-dimensional flows. We developed PetIBM, a code that uses the same
mathematical formulation as cuIBM, to allow solving larger problems on
distributed CPU systems. Since PetIBM and cuIBM implement exactly the
same numerical method, you'd expect that giving the two codes the same
mesh with the same initial conditions will result in the same solution
(within floating-point error). Not so fast! We rely on external
libraries to solve sparse linear systems of equations: \emph{Cusp} for
GPU devices and PETSc for distributed CPU systems. It turns out, the
iterative solvers may have differences that affect the final solution.

When repeating our previous simulations of the aerodynamics of a snake
cross-section with PetIBM, the solutions do not always match those
computed with cuIBM. At a Reynolds number of 1000, both the
time-averaged lift and drag coefficients match. But at Reynolds equal to
2000, average lift and drag match up to 30 degrees angle-of-attack, but
not at 35 degrees. That means that we don't see lift enhancement (Figure
8) and the main finding of our previous study is not fully replicated.
Looking at the time evolution of the force coefficients for the
simulation with PetIBM at Re=2000 and 35 degrees angle-of-attack, we see
a marked drop in lift after 35 time units (top graph in Figure 9). What
is different in the two codes? Apart from using different linear algebra
libraries, they run on different hardware. Leaving hardware aside for
now, let's focus on the iterative solvers. Both \emph{Cusp} and PETSc
use the same convergence criterion. This is not always the case, and
needs to be checked! We're also not using the same iterative solver with
each library. The cuIBM runs (with \emph{Cusp}) used an algebraic
multigrid preconditioner and conjugate gradient (CG) solver for the
modified-Poisson equation. With PETSc, the CG solver crashed because of
an indefinite preconditioner (having both positive and negative
eigenvalues), and we had to select a different method: we used a bi-CG
stabilized algorithm (while still using an algebraic multigrid
preconditioner).

Could this difference in linear solvers affect our unsteady fluid-flow
solution? The solutions with both codes match at lower angles of attack
(and lower Reynolds numbers), so what is going on? Because PetIBM and
cuIBM use the same method, we don't need to repeat mesh-convergence
analysis. But we did confirm convergence of the solution with respect to
the exit criterion of the iterative solvers. We checked everything
multiple times. In the process, we did find some small discrepancies.
Even a small bug (or two). We found, for example, that the first set of
runs with PetIBM created a slightly different problem set-up, compared
with our previous study, where the body was shifted by less than one
grid-cell width. Rotating the body to achieve different angles of attack
was made around a different center, in each case (one used the grid
origin at 0,0 while the other used the body center of mass). This tiny
difference does result in a different average lift coefficient (bottom
graph in Figure 9)! The time signal of lift coefficient shows that the
drop we were seeing at around 35 time units now occurs closer to 50 time
units, resulting in a different value for the average taken in a range
between 32 and 64. Again, this range for computing the average is a
choice we made. It covers about ten vortex shedding cycles, which seems
enough to calculate the average if the flow is periodic. What is causing
the drop in lift? Visualizations of the wake vortices (Figure 10) show
that a vortex-merging event occurs in the middle of the wake, changing
the near-wake pattern. The previously aligned positive and negative
vortices are replaced by a wider wake with a single clockwise vortex on
the top side and a vortex dipole on the bottom part. With the change in
wake pattern comes a drop in the lift force.

\textbf{Postmortem}. Although PetIBM implements the same
immersed-boundary method and was developed by the same research group,
we were not able to fully replicate the previous findings. The
aerodynamic lift on a snake section at 35 degrees angle-of-attack is a
consequence of the near-wake vortices providing extra suction on the
upper side of the body. When a vortex merger event changes the wake
pattern, lift drops. Vortex merging is a fundamentally two-dimensional
instability, so we expect that this problem won't trouble us in more
realistic 3D simulations. But it is surprising that small
changes---within the bounds of truncation error, roundoff error and
algebraic errors---can trigger this instability, changing the flow
appreciably. Even when the only difference between two equivalent
simulations is the linear algebra library used, there can be challenges
to reproducibility.

\begin{figure}
\begin{center}
\includegraphics[width=0.9\columnwidth]{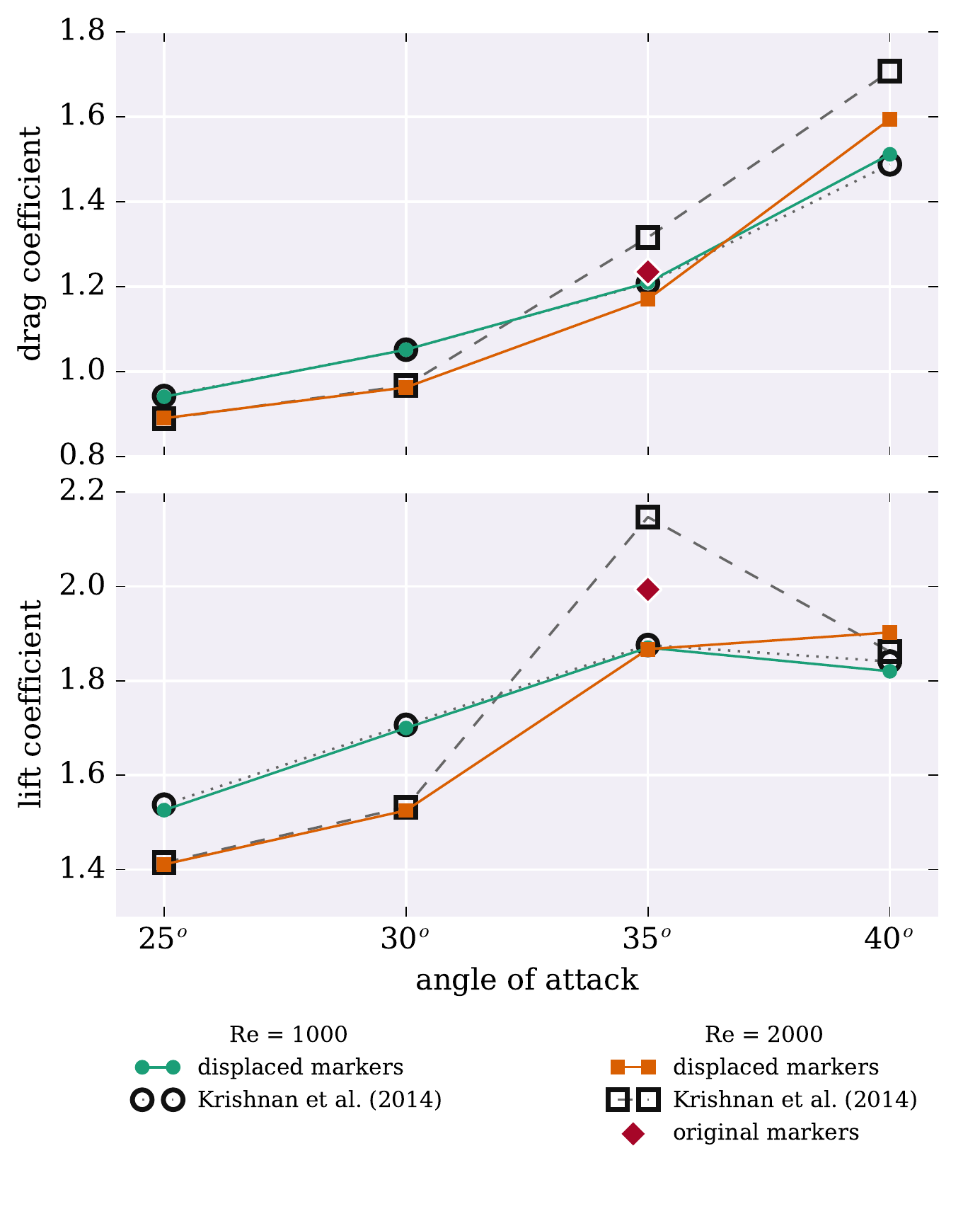}
\caption{{Time-averaged drag (top) and lift (bottom) coefficients as function of
the snake's angle-of-attack and for Reynolds numbers 1000 and 2000 using
the same Eulerian mesh as in our past cuIBM simulations. We show PetIBM
results obtained when the immersed-boundary is rotated around: (1) its
center of mass (green and orange symbols) and (2) the reference origin
(solo red marker). \label{figure8}%
}}
\end{center}
\end{figure}

\begin{figure}[t]
\begin{center}
\includegraphics[width=0.9\columnwidth]{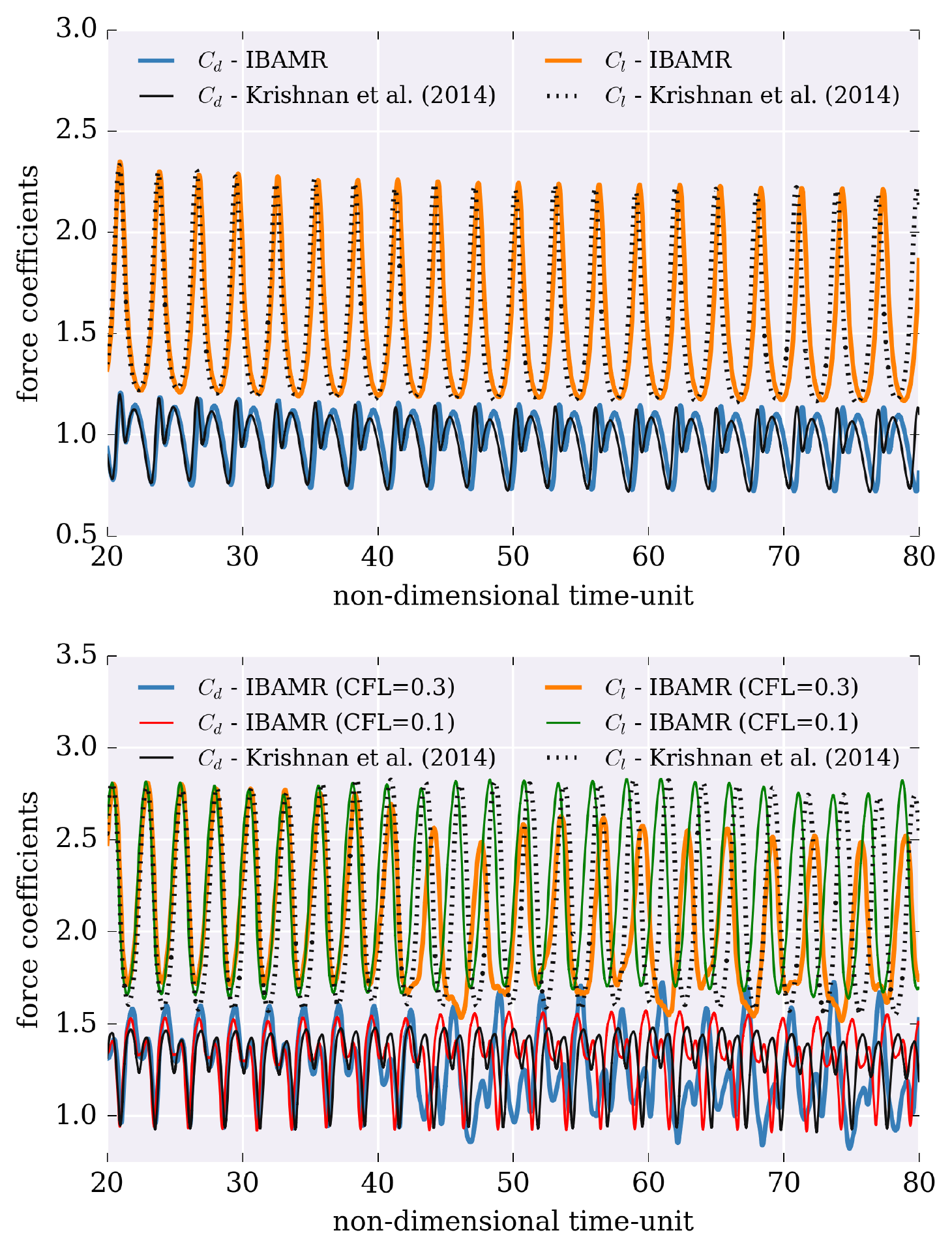}
\caption{{Instantaneous force coefficients for the snake's section with
angle-of-attack 35 degrees and Reynolds number 2000. The top plot
compares the PetIBM results with those reported in our previous study.
The bottom plot shows results with the immersed-boundary being rotated
around the reference origin or around its center of mass (the latter is
slightly shifted compared to our previous study). \label{figure9}%
}}
\end{center}
\end{figure}

\begin{figure}[t]
\begin{center}
\includegraphics[width=0.9\columnwidth]{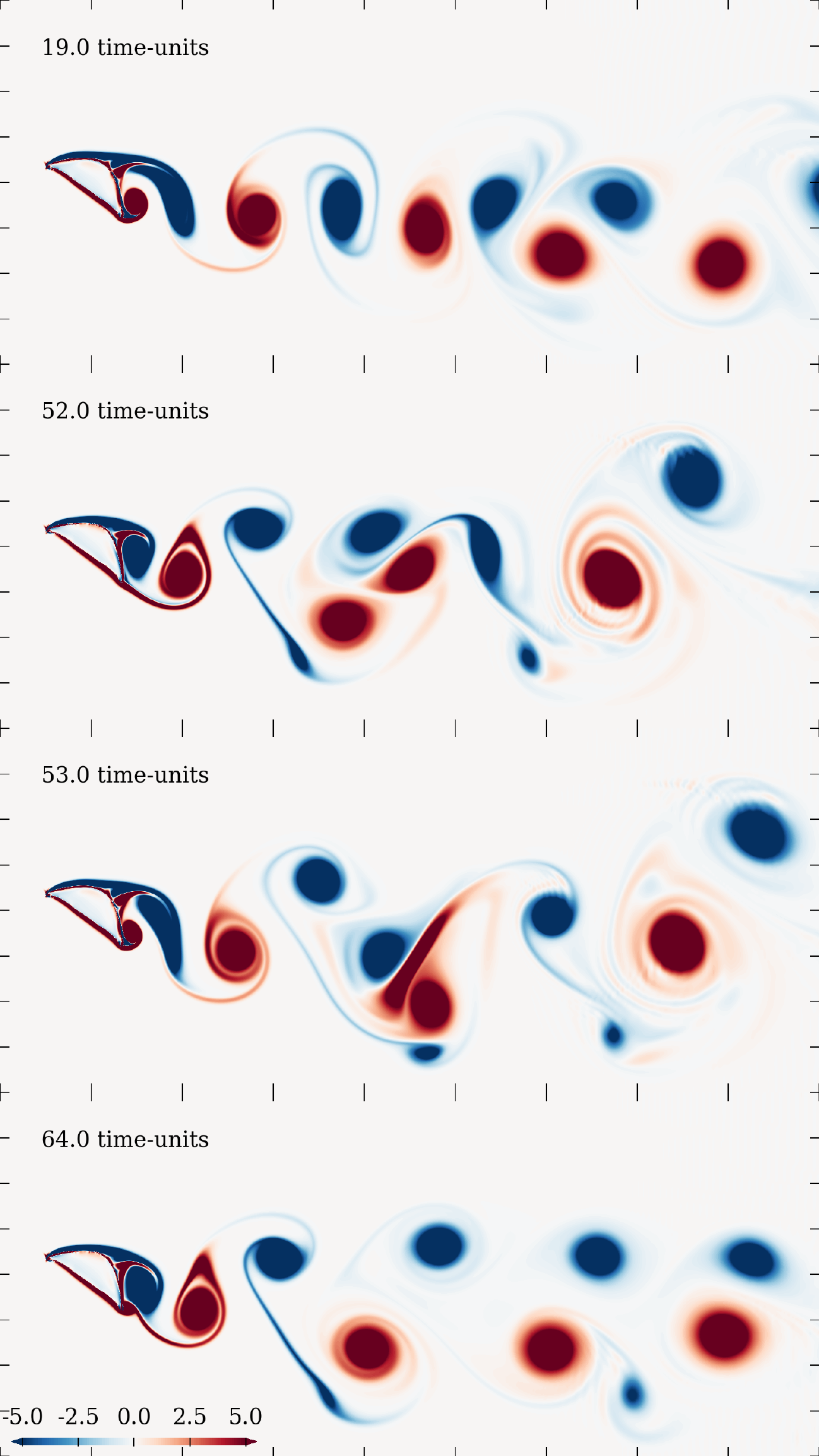}
\caption{{Vorticity field after 19, 52, 53, and 64 time-units of flow-simulation
with PetIBM for a snake's section at angle-of-attack 35 degrees and
Reynolds number 2000. The vortex-merging event is responsible for the
change in the wake signature and the drop in the mean lift coefficient.
\label{figure10}%
}}
\end{center}
\end{figure}

\section*{Story 4: Different versions of your code, external libraries
or even compilers may challenge
reproducibility}\label{story-4-different-versions-of-your-code-external-libraries-or-even-compilers-may-challenge-reproducibility}

In the span of about three years, we ran more than 100 simulations with
OpenFOAM, IBAMR, and PetIBM, encountering about a dozen things that can
go wrong. We replicated our previous scientific finding (enhanced lift
at 35 degrees angle-of-attack for sufficiently large Reynolds number) in
two out of three campaigns. Ironically, the case that did not replicate
our findings was that of our own code re-write. The original code
(cuIBM) and the re-write (PetIBM) use different linear algebra
libraries, and it's unnerving to think this could change our results.
This final story is about what happened when we went back to our
\emph{original} code and tried to reproduce the published findings.

As we mentioned in the opening of this article, we adopted a set of
practices years ago to make our research reproducible. The study
published as ``Lift and wakes of flying snakes'' followed the guidance
of the ``Reproducibility PI Manifesto,'' which includes: (1) code
developed under version control; (2) completed validation and
verification, with report published on Figshare; (3) open data and
figures for the main results of the paper on Figshare; (4) pre-print
made available on arXiv; (5) code released under MIT License; (6) a
Reproducibility statement in the paper. The original work, of course,
confirmed grid independence of the solution: Krishnan et al. (2014) report
differences in the average lift coefficients in the order of 2\% at 35
degrees angle-of-attack and \textless{} 0.1\% at 30 degrees. Naturally,
we expected to be able to reproduce our own results!

The first hurdle we faced is that, three years after we completed our
previous study, we have updated our lab computers: new operating
systems, new GPU devices, new external libraries. The code itself has
been modified to implement new features. Happily, we have version
control. So, we set out to reproduce our results with cuIBM using (1)
the ``same'' old version of the code and (2) the current version. In
both cases, we used identical input parameters (Lagrangian markers to
discretize the geometry, grid parameters, flow conditions, and solver
parameters). But the three-year-old simulations used a version of
\emph{Cusp} (0.3.1) that is no longer compatible with the oldest CUDA
version installed on our machines (5.0). Thus, we adapted ``old'' cuIBM
to be compatible with the oldest version of \emph{Cusp} (0.4.0) that we
can run. The case at angle-of-attack 35 degrees and Reynolds number 2000
now gave an appreciable difference compared with our previous study: the
instantaneous force coefficients start to slowly drop after about 60
time units (Figure 11(c)). Now, this is \emph{really} the same code,
with only a difference in the \emph{version} of the linear algebra
library. Repeating the case with the most current version of cuIBM and
the same version of \emph{Cusp} (0.4.0) leads to the same force signals,
with a slight drop towards the end (Figure 11(d)). And the same is the
case with the current version of cuIBM and a later version of
\emph{Cusp} (0.5.1). The final \emph{findings} in these cases do not
vary from our published work: there is, in fact, lift enhancement at 35
degrees angle-of-attack \ldots{} but the results match only because we
calculate the average lift in a time interval between 32 and 64. Yet,
the flow solution was affected by changing the version of a dependent
library. (The revision history of \emph{Cusp} says that they refactored
the smooth-aggregation solver between the two versions we are using.)
The hardware was also different (a K20 GPU versus a C2070 in our older
study), and the operating system, and the compiler. (Note that we always
run in double precision.) In an iterative linear solver, any of these
things could be related to lack of floating-point reproducibility. And
in unsteady fluid dynamics, small floating-point differences can add up
over thousands of time steps to eventually trigger a flow instability
(like vortex merging).

\textbf{Postmortem}. Making research codes open source is not enough for
reproducibility: we must be meticulous in documenting every dependency
and the versions used. Unfortunately, some of those dependencies will
get stale over time, and might cease to be available or usable. Your
application code may give the same answer with a different version of an
external library, or it may not. In the case of unsteady fluid dynamics,
the nonlinear nature of the equations combined with numerical
non-reproducibility of iterative linear solvers (in parallel!) can
change the results.

\begin{figure*}
\begin{center}
\includegraphics[width=0.95\textwidth]{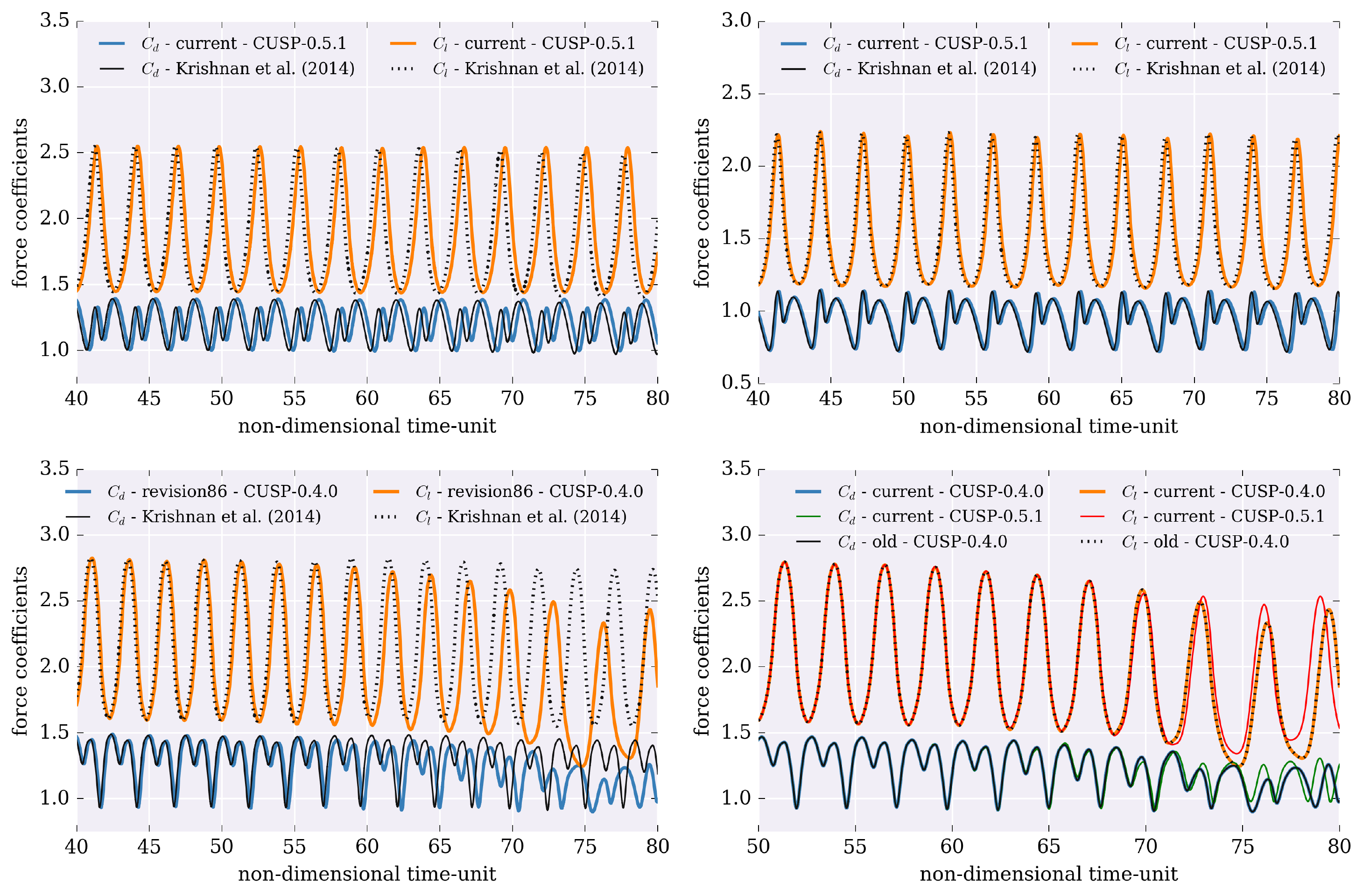}
\caption{{Instantaneous force coefficients on the snake's section at Reynolds
number 1000 and angle 35 degrees (top-left) and Reynolds number 2000 and
angle 30 degrees (top-right). Bottom-left: instantaneous force
coefficients at Reynolds number 2000 and angle-of-attack 35 degrees
running the same version of cuIBM (adapted to CUSP-0.4.0) than the one
used for our previous study. Bottom-right: drop of the mean force
coefficients observed over the end of the simulation using two versions
of cuIBM (``current'' and ``old'') with different CUSP versions (0.4.0
and 0.5.1). \label{figure11}%
}}
\end{center}
\end{figure*}

\begin{figure*}
    \colorbox{lightgray}{
     \begin{minipage}[c]{\textwidth}
      \bigskip
      \subsection*{Definition of reproducible research: }
\footnotesize \sffamily{
 The literature is rife with confused and sometimes contradictory
meanings for reproducible research, reproducibility, replicability,
repetition, etc. It is thus worth clarifying what we mean by these
terms. The phrase ``reproducible research'' in reference to
computational studies that can be reproduced by other scientists was
introduced by geophysics professor Jon Claerbout in the 1990s. An early
volume of CiSE published an article about the reproducible-research
environment they created in his group at Stanford.\cite{schwab2000}
Their goal was complete documentation of scientific computations, in
such a way that a reader can reproduce all the results and figures in a
paper using the author-provided computer programs and raw data. This
ability requires open data and open-source software, and for this reason
the \emph{reproducibility} movement is closely linked with the
\emph{open science} movement. In following years, the term
\emph{replication} was distinctly adopted to refer to an independent
study, re-running experiments to generate new data that, when analyzed,
leads to the same findings. We follow this convention, clarified more
recently in an article in \emph{Science}.\cite{peng2011}

 \medskip
 
\textbf{Reproducible research}--- Authors of a study provide their code
and data, allowing readers to inspect and re-run the analysis to
recreate the figures in the paper. Reproducible research makes
replication easier.\\
\textbf{Replication}--- An independent study that generates new data,
using similar or different methods, and analyzes it to arrive at the
same scientific findings as the original study. 

 \medskip
 
CiSE has dedicated two theme issues to reproducibility: January/
February 2009 and July/August 2012.
      }
      \vspace{0.2cm}
     \end{minipage}}
\end{figure*}

\section*{Lessons learned}\label{lessons-learned}

Reproducibility and replication of studies are essential for the
progress of science, and much of science today advances via computation.
We use computer simulations to create new knowledge. How can we certify
that this new knowledge is justified, that there is enough evidence to
support it? The truth is computational science and engineering lacks an
accepted standard of evidence. Some previous efforts in CFD have sought
to compare results from multiple codes: e.g., Dimonte et al.\ (2004)\cite{Dimonte_2004}
report results for Rayleigh-Taylor instability using seven different
codes (from five institutions), while the AIAA drag-prediction workshop\cite{Levy_2014}
and the high-lift-prediction workshop\cite{Rumsey_2015}
have been collecting results for years using a
variety of commercial and non-commercial (mostly closed) software.
However, these efforts don't have a specific goal of replicating a
published finding nor are they concerned with reproducible workflows. We
label computational research \emph{reproducible} when authors provide
all the necessary data and the computer code to run the analysis again,
re-creating the results. But what data are necessary? We found that
open-source code and open data sets are a minimal requirement.
Exhaustive documentation during the process of computational research is
key. This includes documenting all failures. Current publication custom
is biased towards positive results.\cite{Ioannidis_2005} The CFD
community does not have a habit of communicating negative results; one
rare example is the analysis of Godunov methods and its failures by
Quirk (1997).\cite{quirk1997} In the case of IBAMR, negative results with points
only on the boundary are not among the examples provided: the situation
may be obvious to the authors, but not to the users. We learned how
important the computational mesh and the boundary conditions can be. A
reproducible computational paper should include the actual meshes used
in the study (or a deterministic mesh-generation code) and careful
reporting of boundary conditions. This is rarely (if ever!) the case. We
learned that in addition to developing our code under version control,
we need to carefully record the versions used for all dependencies. In
practice, such careful documentation is feasible only with a fully
automated workflow: launching simulations via running scripts, storing
command-line arguments for every run, capturing complete environment
settings. Post-processing and visualization ideally should also be
scripted, avoiding software GUIs for manipulation of images. New tools
have emerged to help reproducible workflows; for example, Docker
containers to capture the full state of the operating system,
application software, and dependencies.

We learned that highly unsteady fluid dynamics is a particularly tough
application for reproducibility. The Navier-Stokes equations are
nonlinear and can exhibit chaotic behavior under certain conditions
(e.g., geometry, Reynolds number, external forcing). Some flow
situations are subject to instabilities, like vortex merging in two
dimensions and other vortex instabilities in 3D. In any application that
has sufficient complexity, we should repeat simulations checking how
robust they are to small variations. And report negative results!
Understandably, long 3D simulations that take huge computational
resources may not be feasible to repeat. We should continue the
conversation about what it means to do reproducible research in
high-performance computing (HPC) scenarios. When large simulations run
on specific hardware with one-off compute allocations, they are unlikely
to be reproduced. In this case, it is even more important that
researchers advance towards these HPC applications on a solid
progression of fully reproducible research at the smaller scales.

Computational science and engineering makes ubiquitous use of linear
algebra libraries like PETSc, Hypre, Trilinos and many others. Rarely do
we consider that using different libraries might produce different
results. But that is the case. Sparse iterative solvers use various
definitions of the \emph{tolerance} criterion to exit the iterations,
for example. The very definition of \emph{residual} could be different.
This means that even when we set the same value of the tolerance,
different libraries may declare convergence differently! This poses a
challenge to reproducibility, even if the application is not sensitive
to algebraic error. The situation is aggravated by parallel execution.
Global operations on distributed vectors and matrices are subject to
rounding errors that can accumulate to introduce uncertainty in the
results.

We are recommending more rigorous standards of evidence for
computational science and engineering, but the reality is that most CFD
papers are not even accompanied by a release of code and data. The
reasons for this are varied: historical, commercial interests, academic
incentives, time efficiency, export controls, etc. The origins of CFD in
the Los Alamos Laboratory in the 1940s was secret research, and when
computer code was stored in large boxes of punched cards or big rolls of
magnetic tape, there was hardly a way to ``share'' it.\cite{metropolis1982}
The 1970s saw the birth of commercial CFD, when
university professors and their students founded companies funded under
the US government's SBIR program. It's not unreasonable to speculate
that the potential for commercial exploitation was a deterrent for
open-source release of CFD codes for a long time. It is only in the last
15 years or so that open-source CFD codes have become available. But the
CFD literature became entrenched in the habit of publishing results
without making available the code that generated those results. And now,
we face the clash between the academic incentive system and the fact
that reproducible research takes a substantial amount of time and
effort. This campaign to replicate our previous results taught us many
lessons on how to improve our reproducibility practices, and we are
committed to maintaining this high standard. We will continue to share
our experiences.

\subsection*{Supplementary materials}

We provide supplementary materials in the GitHub repository for this
paper, including: (1) all input files to generate the runs reported in
the paper; (2) Jupyter notebooks for all the simulations, detailing
every condition (dependencies, compilation, mesh, boundary conditions,
solver parameters, command-line options); (3) Python codes needed to
recreate all the figures included in the paper. In addition, we
separately report our efforts to assess independence of the solution
with respect to grid spacing, time increment and iterative tolerance
(with each code).

Our codes are available for unrestricted use, under the MIT license; to
obtain the codes and run the tests in this paper, the reader may follow
instructions on \url{https://github.com/barbagroup/snake-repro}.

\bigskip

\section*{Acknowledgements}

{\sf \emph{\small We're grateful for the support from the US National Science Foundation
for grant number NSF OCI-0946441 and from NVIDIA Corp. for equipment
donations under the CUDA Fellows Program. LAB would like to acknowledge
the hospitality of the Berkeley Institute of Data Science (BIDS), where
this paper was written.
}

\bibliographystyle{unsrt}
\bibliography{snake-repro}

\vspace{1cm}

\small
{\sf

\noindent \textbf{Olivier Mesnard} is a doctoral student at the George Washington University. 
He has an Engineering degree from Ecole Sup{\'e}rieur de M{\'e}canique (SUPMECA) in Paris, 
and a Master's degree in Sciences from Universit{\'e} Pierre et Marie Curie, also in Paris.
His research interests include computational fluid dynamics and immersed-boundary methods 
with application to animal locomotion.

\bigskip

\noindent\textbf{Lorena A. Barba} is Associate Professor of Mechanical and Aerospace Engineering at the George Washington University.
She obtained her PhD in Aeronautics from the California Institute of Technology.
Her research interests include computational fluid dynamics, especially particle methods for fluid simulation and immersed boundary methods; fundamental and applied aspects of fluid dynamics, especially flows dominated by vorticity dynamics; the fast multipole method and applications; and scientific computing on GPU architecture. 
She received the Amelia Earhart Fellowship, a First Grant award from the UK Engineering and Physical Sciences (EPSRC), the National Science Foundation Early CAREER award, and was named CUDA Fellow by NVIDIA, Corp.
}

\end{document}